# Reducing number of gates in quantum random walk search algorithm via modification of coin operators

**Hristo Tonchev** [1,2]*

**Petar Danev** [2]

[1] Institute of Solid State Physics, Bulgarian Academy of Sciences,72 Tzarigradsko Chaussée, 1784 Sofia, Bulgaria

[2] Institute for Nuclear Research and Nuclear Energy, Bulgarian Academy of Sciences,72 Tzarigradsko Chaussée, 1784 Sofia, Bulgaria

* Correspondence: htonchev@issp.bas.bg;

**Abstract:** This paper examines a way to simplify the circuit of quantum random walk search algorithm, when the traversing coin is constructed by both generalized Householder reflection and an additional phase multiplier. If an appropriate relation between corresponding parameters is realized, our algorithm becomes more robust to deviations in the phases. In this modification marking coin is not needed, and all advantages from above mentioned optimization to the stability, are preserved. It is shown explicitly how to construct such walk coin in order to obtain more robust quantum algorithm.

**Keywords:** Quantum algorithms, Quantum Random Walk, Quantum Search, Generalized Householder Reflection

1. Introduction

A quantum analogue of classical random walk [1] was first introduced by Y. Aharonov at. al [2]. At the beginning it was used to traverse simple structures like line [3] and circle [3]. Later - for much more complex objects like square [4] and hexagonal grids [5], torus [4] and hypercube [6]. It's fast hitting time and possibility to be used to traverse structures with arbitrary topology makes this algorithm useful subroutine in variety of other applications like quantum cryptography and other quantum algorithms. A few examples are: image encryption [7] and quantum hash function than can be used in quantum cryptographic protocols [8], quantum algorithms for calculating Boolean formulas [9], quantum algorithm for finding triangles [10] and quantum random walk search (QRWS) [11].

Quantum random walk search was developed by Shenvi at. al. This algorithm is probabilistic. Together with Grover's search algorithm [12], both are quadratically faster than classical search. However, they excel in different tasks. In contrast to Grover search, suited only for linear database, QRWS can be used to search in database with arbitrary topology including: simplex [13], hypercube [14], fractal lattice [15] and hypercubic lattice [16].

Many modifications of quantum walk search exists, some of them are general and can be used on different topologies like increasing the number of steps per oracle query [13] and QRWS with more than one solution [17] [18]. Other are suited to one particular system like increasing the probability to find solution for hypercube [14].

Householder reflection is widely used in quantum information. Quantum gates can be efficiently created by using Householder reflections and phase gates [19]. Decomposition to reflections is quadratically faster, compared to the other popular method - decomposition to Givens rotations [20]. Generalized householder reflection, in both cases of qubits and qudits, can be efficiently implemented in some physical systems like linear ion traps [21] and photonic quantum computer [22].

In our previous work [23] we proposed a method for construction of QRWS walk coin for Hypercube by using generalized Householder reflection and a phase multiplier. We show that QRWS is more robust to variations in coin's phase parameters, if an appropriate functional dependence between the additional phase and the one involved in the reflection is maintained. We demonstrate that there is an increase in the robustness for linear relation between the phases and that the stability improves further if the physical implementation allows nonlinear relation to be established. In [24] we apply the same method for qudit walk coin.

In this work we show how the number of gates needed to construct QRWS on hypercube can be reduced while maintaining robustness of the algorithm. This article is organized as follows:

In Sec. [2] we review our stability modification of QRWS algorithm. In Subsec. [2.1] briefly discuss QRWS algorithm and its quantum circuit. Subsec. [2.2] shows how walk coin can be constructed by using Generalized Householder reflection and phase multiplier. Monte Carlo simulations of QRWS algorithm is are given in Subsec. [2.3]. The final Subsec. [2.4] is dedicated to discussion of the optimized algorithm's stability. In Sec. [3] we show new more general modification of quantum random walk search algorithm. The analytical study of this construction is supported with Monte Carlo simulations. We prove that the high robustness of the algorithm is preserved. Next in Sec. [4], we show how this modification can be used to remove the need to build a marking coin, as it becomes just the identity operator. In the conclusion we summarize our results.

### 2.1. Quantum random walk search on hypercube

QRWS algorithm is quantum algorithm designed for searching in unordered database and can be used on graphs with arbitrary topology. Quantum circuit of QRWS algorithm is showed on FIG. 1:

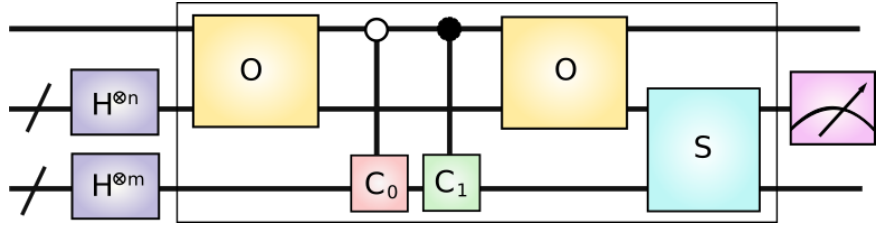

FIG. 1: Quantum circuit for QRWS algorithm. It contains the following quantum gates: Oracle O, Hadamard gate H, Shift operator S, traversing and marking coins $C_0$ and $C_1$

Quantum algorithm begins with applying Hadamard gate to the register of nodes (consisting $2^n$ states) and register of edges (consisting of $n = 2^m$ states) to put them in equal weight superposition. Next, the iteration of QRWS should be applied a certain number of times k. The iteration itself consists of the following sequence of operators: 1) Oracle $O$ on the control and node register. Oracle recognizes the nodes that are solutions and marks them. 2) Traversing $C_0$ and marking coins $C_1$ should be applied depending on whether the node is marked by oracle or not. Traversing coin gives the probabilities to go at each direction. 3) Oracle is applied again. The iteration ends with Shift operator $S$ that defines the topology of the walked object. At the end the register of nodes is measured. If right solution is not found, the algorithm is repeated.

When the graph constructed by all nodes and edges is a hypercube the probability to find solution of the original QRWS (with Grover coin used as traversing coin) is approximately $1/2 - O(1/2^m)$ [11]. In the case when the algorithm has only one solution, number of iterations $k$ can be calculated as:

$$k = \left\lceil \frac{\pi}{2}\sqrt{2^{m-1}} \right\rceil \qquad (1)$$

### 2.2. Matrix elements of the Walk coin and Generalized Householder Reflection

In the case when the probability to go in each direction is the same (regular graph), walking coin have the matrix representation with one value for all matrix elements on the main diagonal and another value for all other elements:

$$C_0 = \begin{pmatrix} a & b & \cdots & b \\ b & a & \cdots & b \\ \vdots & \vdots & \ddots & \vdots \\ b & b & \cdots & a \end{pmatrix} \qquad (2)$$

Matrix representation of $C_0$ is complex and have dimension $r$. If the coin is constructed of qubits, $r$ is power of two. However, the coin can be also made by arbitrary number of qudits with dimension $m$, so $r$ will be power of $m$. In our work the coin consists of one qudit with arbitrary dimension $r = m$.

Such walk coin can be obtained by using generalized Householder reflection:

$$M(\phi, \chi) = I - (1 - e^{i\phi})|\chi\rangle\langle\chi| \qquad (3)$$

Here, $\phi$ is a phase and $|\chi\rangle = \frac{1}{\sqrt{m}}\sum_{i=0}^{m-1}|i\rangle$ – an equal weight superposition of coin register's basis vectors $|i\rangle$. In our previous work [23], we construct the traversing coin $C_0$ by using only one generalized Householder reflection and additional phase multiplier $e^{i\zeta}$:

$$C_0(\phi, \chi, \zeta) = e^{i\zeta} M(\phi, \chi) = e^{i\zeta}(I - (1 - e^{i\phi})|\chi\rangle\langle\chi|) \qquad (4)$$

Maximum probability to find solution depends on the coin used. As an example, Grover coin can be obtained by angles $\phi = \zeta = \pi$. For large enough node register it gives probability to find solution close to ½. Another example is identity operator if $\phi = \zeta = 0$. In this case there is no random walk. Maximum probability to find solution with the modified walk coin (4) now has dependance on both phases $\phi$ and $\zeta$, and can be written as:

$$p(\phi, \zeta, m) = \frac{1}{2} - f(\phi, \zeta) \times O\left(\frac{1}{2^m}\right) \qquad (5)$$

### 2.3. Monte Carlo simulations of QRWS

In our previous works [23], [24] we made Monte Carlo simulations of the QRWS with walk coin (4) constructed by one qudit with size m. For random values of $\phi, \zeta \in [0, 2\pi]$, the resulted probability to find solution is $p(\phi, \zeta, m)$.

Monte Carlo simulation examples of QRWS algorithm for modified traversing coin (4) with size 4 and 7 are shown on FIG. 4. The higher probability to find solution corresponds to darker color and lower probability - to lighter. From the figures it is obvious that with the increase of the coin size the high probability area's width decreases.

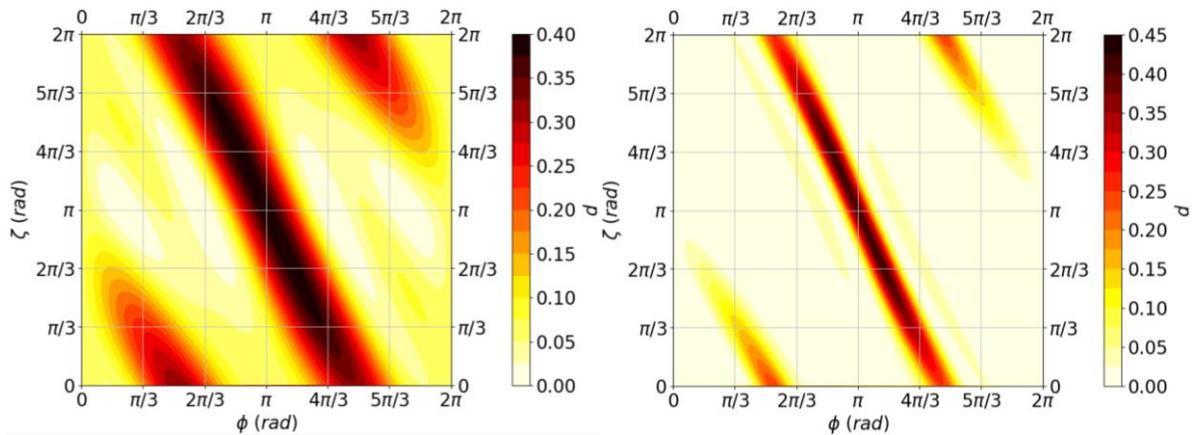

FIG. 2: Probability to find solution for coin size m=4 (left) and m=7 (right). Horizontal axis corresponds to Householder Reflection angle $\phi$ and vertical - to $\zeta$. Lighter color represents lower probability to find solution and darker - higher probability.

## 2.4. Robustness of QRWS with Householder coin

The specific form of the functional dependence $p(\phi, \zeta, m)$ (shown on FIG. 4 for coin size m = 4 and 7) justifies us to look for relations $\zeta = \zeta(\phi, m = const)$ leading to high probability $p(\phi, \zeta(\phi), m = const) \equiv p(\phi)$, $\phi \in [\phi_{max} - \varepsilon, \phi_{max} - \varepsilon]$ for the largest possible interval $2\varepsilon$. Here $\phi_{max}$ is the value of $\phi$ where the maximal probability to find solution is achieved for fixed m.

We have searched for optimal function $\zeta(\phi)$ by using Monte Carlo simulations. The best results were obtained [23] with:

$$\zeta = -2\phi + \pi + \alpha \sin(2\phi). \tag{6}$$

The optimal parameters $\alpha = \alpha_{ML}$ for different coin register size m, were found by machine learning methods. The values of $\alpha_{ML}$ for different coin sizes can be found in [24], [23].

On FIG. 3 are shown examples of the simulation results for $p(\phi)$ with walk coins generated by Generalized Householder reflection and phase $\zeta = \zeta(\phi)$ for coin size m = 4(left), m = 7(right). The teal dashed line corresponds to the unmodified quantum walk algorithm with $\zeta = \pi$. The red dot-dashed one is obtained if relation (6) with $\alpha = 0$ is used. The purple dotted and the solid green lines correspond to (6) where $\alpha = -1/(2\pi)$ and $\alpha = \alpha_{ML}$ respectively. From both figures it is evident that the high probability plateau is much wider if the walk coin in QRWS is constructed according to Eq. (6). The most stable algorithms are obtained when nonlinear relations between the walk coin phases are used. For example, $\alpha = -1/(2\pi)$ always gives better results than the linear function. The above-mentioned modification of QRWS makes it more robust to deviations of the parameter $\phi$ from its optimal value, and we expect - more stable experimental implementation.

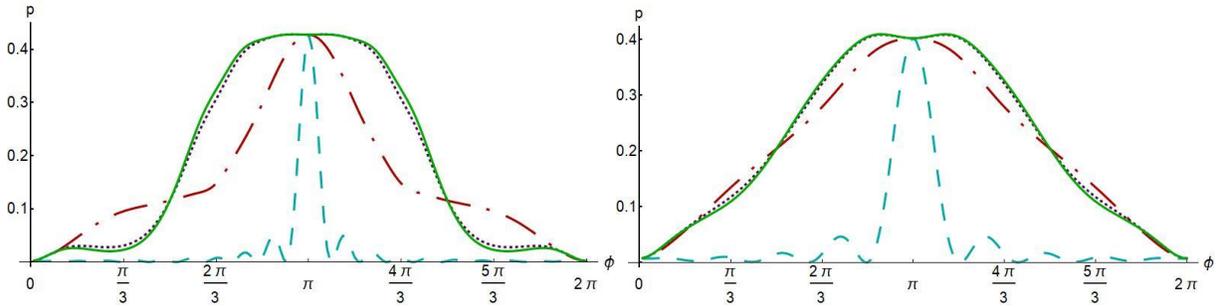

FIG. 3: Probability to find solution of QRWS algorithm $p(\phi, \zeta(\phi), m = const)$ for different coin phases' relations $\zeta(\phi)$. m = 4(left), m = 7(right). The teal dashed line corresponds to the unmodified quantum walk algorithm with $\zeta = \pi$. The red dot-dashed – to relation (6) with $\alpha = 0$ is used. The purple dotted and the solid green lines correspond to (6) where $\alpha = -1/(2\pi)$ and $\alpha = \alpha_{ML}$ respectively.

## 3. QRWS on hypercube with modification of both coins

In our previous works [23], [24] we have studied only the case when the walk coin is modified. Here we investigate modification of the marking coin. Our goal is to reduce the number of quantum gates needed while preserving the high robustness of the algorithm.

Here, we use the same walk coin defined in Eq. (4). The original QRWS algorithm use the operator

$$C_1 = -I \tag{7}$$

for marking coin. Here will be used the identity operator multiplied by a new phase $\omega$:

$$C'_1 = -e^{i\omega}I \tag{8}$$

Quantum circuit is the same as in the original algorithm (see FIG. 1). The phase factors in front of walk and mark coins cause constructive and destructive interferences. The phase difference between them is crucial as shown in the next paragraphs.

Let $\mathbb{C}_0 = Diag(I, C_0)$ is controlled gate that apply coin $C_0$ when element is not solution and similarly $\mathbb{C}_1 = Diag(C_1, I)$ when element is solution. The corresponding controlled gate with modified marking coin $C'_1$ will be $\mathbb{C}'_1 = Diag(C'_1, I)$. The symbol $I$ is an identity operator with dimension equal to that of $C_0$. Consequentially applying both operators can be described with the following matrix form in case when the marking coin is given by Eq. (7):

$$\mathbb{C}(\phi, \chi, \zeta) = \mathbb{C}_0 \mathbb{C}_1 = Diag(e^{i\zeta}M(\phi, \chi), -I) \tag{9}$$

And when $C'_1$ (Eq. (8)) is used:

$$\mathbb{C}'(\phi, \chi, \zeta, \omega) = \mathbb{C}_0 \mathbb{C}'_1 = Diag\left(e^{i\zeta}M(\phi, \chi), e^{i\omega}(-I)\right) \tag{10}$$

$$= e^{i\omega} Diag(e^{i(\zeta-\omega)}M(\phi, \chi), -I)$$

We can rewrite $\mathbb{C}'$ as:

$$\mathbb{C}'(\phi, \chi, \zeta, \omega) = e^{i\omega}\mathbb{C}(\phi, \chi, \zeta - \omega) \tag{11}$$

It is trivial to see that $e^{i\omega}$ is global phase for each iteration of the algorithm and can be neglected. Then, if we substitute $\Delta = \zeta - \omega$, the controlled gates $\mathbb{C}'(\phi, \chi, \Delta, \omega) \rightarrow \mathbb{C}'(\phi, \chi, \Delta)$ and $\mathbb{C}(\phi, \chi, \Delta)$ lead to the same results from the execution of the algorithm and their probability distributions are the same. So, the phase difference between coins $\Delta$ is important and not the specific values of $\omega$ and $\zeta$. If the phases are $\omega = 0$ and $\phi = \zeta = \pi$, this is the standard quantum random walk search as proposed by [11].

On FIG. 4 are presented the results from Monte Carlo simulation of QRWS algorithm for different values of $\omega$ and different coin size m. The parameters $\phi$ and $\zeta$ did not change their values between figures. Lighter color corresponds to lower probability to find solution, darker - to higher. On the first row are pictures for a coin with size 4, and on the second - for coin size 7. Each column corresponds to different value of $\omega = [0, \frac{\pi}{2}, \pi, \frac{3\pi}{2}]$.

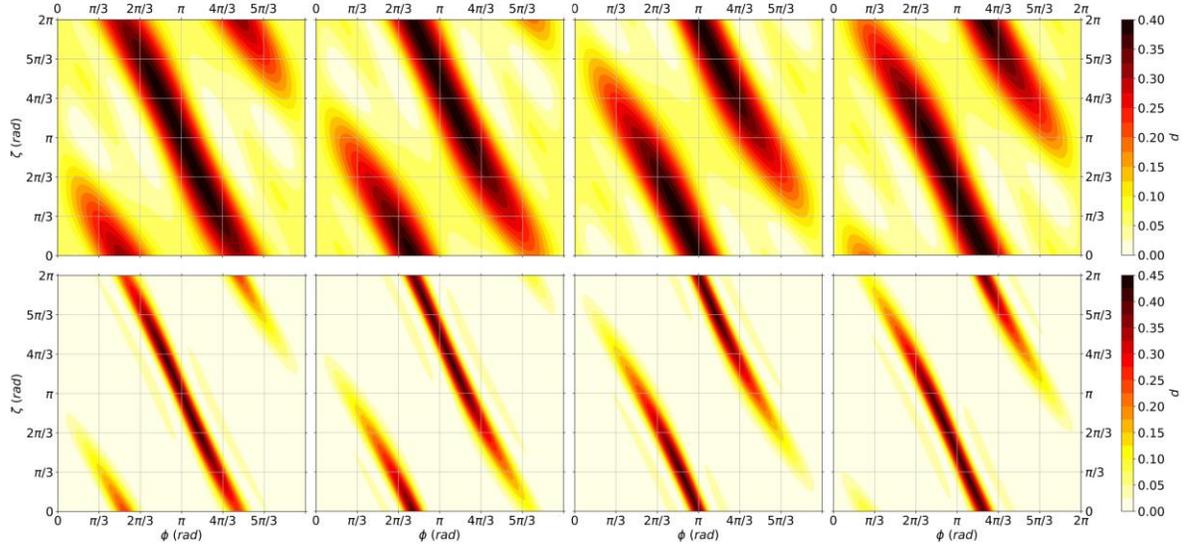

FIG. 4: Monte Carlo simulation of QRWS algorithm for different values of coins' parameters. Traversing coin phases $\phi$ and $\zeta$ are given on $x$ and $y$ axes. Each column (from left to right) corresponds to different value of $\omega = [0, \frac{\pi}{2}, \pi, \frac{3\pi}{2}]$, and first and second rows – to m = 4 and m = 7 respectively. The probability $p(\phi, \zeta(\phi), \omega, m)$ increases from lighter to darker colors.

From the pictures on FIG. 4 it is obvious that the phase difference $\Delta = \zeta - \omega$ leads to translation of the distribution on the $\zeta$ axis. This observation confirms the analytical calculations given above.

With counting the effect of the phase of the marking coin, from Eq. (10) it follows that Eq. (6) changes to:

$$\zeta = -2\phi + \omega + \pi + \alpha \sin(2\phi), \quad \phi \epsilon [0, 2\pi]. \tag{12}$$

The values of $\alpha(m)$ giving the highest robustness of the quantum algorithm depend only on the shape and size of the area with high probability $p(\phi, \zeta(\phi), \omega, m)$, which remains the same as is shown in [24]. Thus, adding a phase factor to the marking coin does not change the values of the parameter $\alpha$.

On FIG. 5 is shown the probability to find solution p of QRWS algorithm with walk coin constructed by Eq. (4) and marking coin by Eq. (8). Different lines correspond to different relations $\zeta(\phi)$: the teal dashed line - to the unmodified quantum walk algorithm with $\zeta = \pi$. The red dot-dashed one is obtained from relation (6) for the left pictures and (12) for the right ones with $\alpha = 0$. The purple dotted and the solid green lines correspond to (6) (or (12) accordingly) with $\alpha = -1/(2\pi)$ and $\alpha = \alpha_{ML}$ respectively. First and second rows correspond to coin size m = 4, 7. For all figures $\omega = \pi/3$. In the pictures on the left the algorithm is simulated with $\zeta$ given by Eq. (6), and in the ones on the right – with $\zeta$ given by Eq. (12) that keeps the phase difference $\Delta$ constant. The simulation results show that if $\Delta$ is changed then the probabilities $p(\phi, \zeta(\phi), \omega, m)$ decrease for all lines representing our high robustness modification of QRWS algorithm [23]. Contrary, if $\zeta$ and $\omega$ are related by Eq. (12) and therefore the magnitude of $\Delta$ is preserved, the modified quantum algorithm retains high stability and high probability to find solution $p(\phi, \Delta, m)$ as seen on FIG. 5 (right). The latter pictures coincide with the ones shown on FIG. 3, where $\omega = 0$ and $\zeta(\phi)$ is given by Eq. (6).

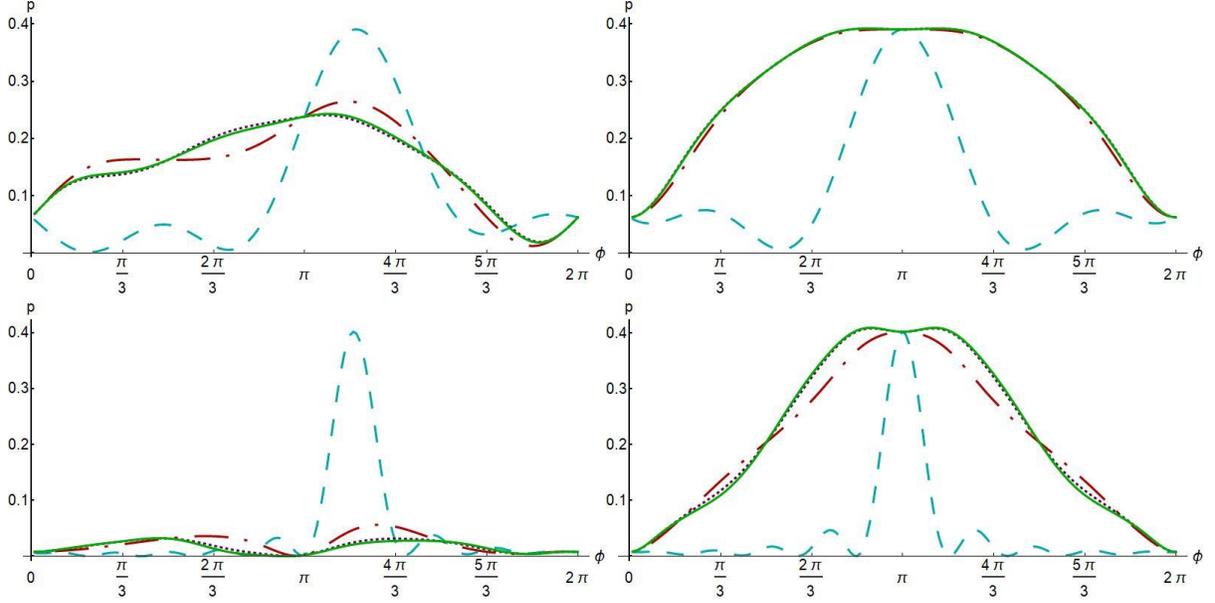

FIG. 5: Probability to find solution of QRWS with walk and marking coins constructed by Eq. (8) and Eq. (4). Different lines correspond to different relations between coin phases: the teal dashed line - to $\zeta = \pi$, the red dot-dashed, the purple dotted and the solid green ones are obtained with relation (6) for left pictures and (12) for the right ones with $\alpha = 0$, $\alpha = -1/(2\pi)$ and $\alpha = \alpha_{ML}$ respectively. First and second rows correspond to coin size m = 4, 7. For all figures $\omega = \pi/3$. In the pictures on the left the algorithm is simulated with $\zeta$ given by Eq. (6), and in the ones on the right – with $\zeta$ given by Eq. (12)

This confirms that $p$ depends not on angles $\zeta$ and $\omega$ as separate variables, but on their difference $\Delta = \zeta - \omega$. This can be used to simplify the quantum circuit for QRWS algorithm. The recipe and the modified quantum circuit will be shown in the next section.

## 4. Alternative Circuit

Based on the considerations made in the above chapters on more general construction of walk and mark coins, we propose new modification of QRWS algorithm that simultaneously simplifies algorithm's quantum circuit and preserves the robustness resulting from the optimization of the walk coin explained in [23]. The most significant improvement is removing the need of mark coin. The quantum scheme of the optimized algorithm is shown on FIG. 6.

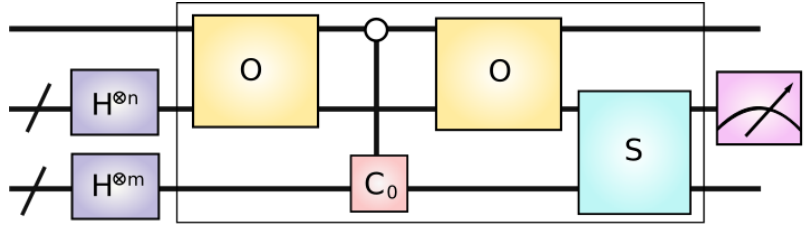

FIG. 6: Alternative quantum circuit of QRWS algorithm on hypercube without use of a marking coin. The circuit consists of Oracle O, Hadamard Gate H, Shift operator S, and traversing coin $C_0$.

In this alternative circuit, the walk coin (7) changes to:

$$C_0(\phi,\chi,\zeta) = e^{i(\zeta-\pi)}\bigl(I - (1 - e^{i\phi})|\chi\rangle\langle\chi|\bigr) \qquad (13)$$

and the marking coin becomes the identity operator $C'_1 = -e^{i\pi}I = I$. Both expressions are derived from Eqs. (8) and (12) when $\omega = \pi$. This allows the algorithm to be built with a walking coin only, without using a marking coin. This improvement will reduce the number of gates needed to make QRWS algorithm. In the particular case when the phases are $\phi = \zeta = \pi$ the walking coin is a minus Grover coin.

The second effect of the above modification is an increased robustness of QRWS to deviations in the coin parameters. All the advantages studied in Section [2.3, 2.4] and in more detail in [24], [23] remain with the following functional relation between coins' phases:

$$\zeta = -2\phi + \alpha\sin(2\phi), \qquad \phi\epsilon[0,2\pi]. \qquad (14)$$

## 5. Conclusion

In this paper an alternative quantum circuit of quantum random walk search on hypercube has been proposed. This modification allows to omit one of the quantum gates, namely the mark coin, if the walk coin is constructed by generalized Householder reflection and additional phase multiplier. Both can be easily implemented on some systems such as the ion traps. This walk coin, as we showed in our previous works, can be used to make the quantum algorithm more robust to inaccuracies in the phases. We show that only the phase difference between both walk and marking coins is important for the quantum random walk search algorithm. The proposed optimization of QRWS retains the same high stability to parameter deviations if a certain functional dependance between them is maintained. The above statements are proven by Monte Carlo simulations and a few selected cases are presented as an example. Our results allow construction of quantum circuit for quantum random walk search on hypercube that contain less quantum gates but remains robust to inaccuracies in the phases used to construct the walk coin.

**Acknowledgments**

The work on this paper was supported by the Bulgarian National Science Fund under Grant KP-06-H58/5 / 12.02.2021.